\documentclass[12pt]{article}
\usepackage[ukrainian,english]{babel}
\usepackage[utf8]{inputenc}
\usepackage{doc}
\usepackage{amsmath}
\usepackage{amsfonts}   
\usepackage{cite} 
\usepackage{fullpage}                           
\usepackage[lmargin=0.6in,rmargin=0.6in,tmargin=0.6in,headsep=.2in]{geometry}
\usepackage{graphicx}
\graphicspath{{Fig/}}
\usepackage{forloop}
\usepackage{etoolbox}
\usepackage{calc}
\usepackage{wrapfig,lipsum,booktabs} 
\usepackage{xtab} 
\usepackage[dvipsnames]{xcolor}
\usepackage[normalem]{ulem}
\usepackage{sectsty}
\usepackage{amsmath,amsfonts,amssymb,amsthm,epsfig,epstopdf,url,array}
\usepackage[retainorgcmds]{IEEEtrantools}
\usepackage{makeidx,epsfig,lscape}
\usepackage{xcolor,pict2e}
\usepackage{upgreek}
\usepackage{pgfplots}
\usepackage{float} 
\usepackage[sloped]{fourier}
\usepackage{comment}

\makeatother
\usepackage{makeidx}
\makeindex
\makeatletter
\renewcommand{\@biblabel}[1]{#1.}

\title{China and G7 in the Current Context of the World Trading}
\author{N.S. Gonchar, O.P. Dovzhyk, A.S. Zhokhin, W.H. Kozyrski, A.P. Makhort\\
	Bogolyubov Institute for Theoretical Physics of NAS of Ukraine.}
\date{}

\begin{document}

\maketitle


\noindent 

\noindent 

\noindent 

\noindent 

\noindent 

\noindent 

\noindent 

\begin{abstract}

The paper analyses trade between the most developed economies of the world. The analysis is based on the previously proposed model of international trade. This model of international trade is based on the theory of general economic equilibrium. The demand for goods in this model is built on the import of goods by each of the countries participating in the trade. The structure of supply of goods in this model is determined by the structure of exports of each country. It is proved that in such a model, given a certain structure of supply and demand, there exists a so-called ideal equilibrium state in which the trade balance of each country is zero. Under certain conditions on the structure of supply and demand, there is an equilibrium state in which each country have a strictly positive trade balance. Among the equilibrium states under a certain structure of supply and demand, there are some that differ from the ones described above. Such states are characterized by the fact that there is an inequitable distribution of income between the participants in the trade. Such states are called degenerate. In this paper, based on the previously proposed model of international trade, an analysis of the dynamics of international trade of 8 of the world's most developed economies is made. It is shown that trade between these countries was not in a state of economic equilibrium. The found relative equilibrium price vector turned out to be very degenerate, which indicates the unequal exchange of goods on the market of the 8 studied countries. An analysis of the dynamics of supply to the market of the world's most developed economies showed an increase in China's share. The same applies to the share of demand.

\end{abstract}


\noindent 

\section{Introduction}


This article is a continuation of works [1-7], in which a model of international trade is proposed and a mathematical method of its research is developed.This became possible thanks to a new approach to the description of the economy proposed in [2].The article [1] provides an analysis of the international trade of the G20 countries.In it, we introduce the concept of the ideal state of equilibrium in world trade. As a rule, the world trade is not being in an ideal state of equilibrium, it deviates from it. The deviation can be due to both the state of each of the economies and the tariff restrictions that each of the countries implements to protect against the flow of goods from the outside.This state of equilibrium is usually violated through a monopoly in the supply of a certain type of goods, oligopoly and cartel agreements. The result of the so-called chaotic tariff war could be a shift in the world economy towards a state of recession. The actual equilibrium state can be significantly different from the ideal equilibrium state. The study of such equilibrium states is an urgent problem in order to prevent the negative development of international trade.

Section 2 contains a general formulation of the problem.

Section 3 shows that the trade between the world's most developed economies was not in a state of economic equilibrium. The relative equilibrium price vector turned out to be highly degenerate.Section 4 presents the dynamics of changes in market share of the 8 most developed economies of the world.The problem of restructuring international trade on fair principles that would ensure sustainable development is discussed in the articles [8-13].

\section{Model Formulation and Statement of the Problem}

This paper is an application of the economy description concept elaborated in [2] to model the international trade between $M$ countries.

Every set of goods we describe by a vector $x=(x_{1} ,\ldots ,x_{n} ),$ where $x_{i} $ is a quantity of units of the $i$-th goods, $e_{i} $ is a unit of its measurement, $x_{i} e_{i} $ is the natural quantity of the goods. If $p_{i} $ is a unit price for the goods $e_{i} ,$$i=\overline{1,n},$ then $p=(p_{1} ,\ldots ,p_{n} )$ is the price vector corresponding to the vector of goods $(e_{1} ,\ldots ,e_{n} ).$ The price of goods vector $x=(x_{1} ,\ldots ,x_{n} )$ is $\left\langle p,x\right\rangle =\sum _{i=1}^{n} p_{i} x_{i} .$ The set of possible goods in the considered period of the economy system operation is denoted by $S.$ We assume $S$ is a convex subset of the set $R_{+}^{n} .$ As for what follows only the property of convexity is important, we assume, without loss of generality, that $S$ is a certain $n$-dimensional parallelepiped that can coincide with $R_{+}^{n} .$ Thus, we assume that in the economy system the set of possible goods $S$ is a convex subset of the non-negative orthant $R_{+}^{n} $ of $n$-dimensional arithmetic space $R^{n} ,$ the set of possible prices is a certain cone $K_{+}^{n} ,$ contained in $R_{+}^{n} \backslash \{ 0\} ,$ and that can coincide with $R_{+}^{n} \backslash \{ 0\} .$

\textit{Definition 1.A set $K_{+}^{n} \subseteq \bar{R}_{+}^{n} $ is called a nonnegative cone if together with a point $u\in K_{+}^{n} $ the point $tu$ belongs to the set $K_{+}^{n} $ for every real $t>0.$}

Here and further, $R_{+}^{n} \backslash \{ 0\} $ is a cone formed from the nonnegative orthant $R_{+}^{n} $ by ejection of the null vector $\{ 0\} =\{ 0,\ldots ,0\} .$ Further, the cone $R_{+}^{n} \backslash \{ 0\} $ is denoted by $R_{+}^{n} .$

Suppose $M$ countries exchange by $n$ types of goods. Let $i_{kj}^{s} $ be the quantity of import units for $s$-th goods from $j$-th country into $k$-th country and $p_{s} i_{kj}^{s} $ its value, where $p=\{ p_{s} \} _{s=1}^{n} $ is a price vector. Also, let $e_{kj}^{s} $ be a quantity of export units for $s$-th goods from the $k$-th country into the $j$-th one and $p_{s} e_{kj}^{s} $ is its value. Having these entities, let us introduce the demand $||c_{sk}^{1} ||_{s=1,k=1}^{n,M} ,$ and supply matrices $||b_{sk}^{1} ||_{s=1,k=1}^{n,M} ,$
\begin{equation} \label{GrindEQ__1_} 
c_{sk}^{1} =\sum _{j=1}^{M} i_{kj}^{s} ,\quad b_{sk}^{1} =\sum _{j=1}^{M} e_{kj}^{s} ,\quad s=\overline{1,n},\quad k=\overline{1,M}, 
\end{equation} 

and the supply vector $\psi ^{1} =\{ \psi _{s}^{1} \} _{s=1}^{n} ,$ where
\begin{equation} \label{GrindEQ__2_} 
\psi _{s}^{1} =\sum _{k=1}^{M} b_{sk}^{1} =\sum _{k,j=1}^{M} e_{kj}^{s} ,\quad s=\overline{1,n}. 
\end{equation} 

The income of the $k$-th country from its export is
\begin{equation} \label{GrindEQ__3_} 
D_{k} (p)=\sum _{s=1}^{n} p_{s} b_{sk}^{1} =\sum _{j=1}^{M} \sum _{s=1}^{n} e_{kj}^{s} p_{s} . 
\end{equation} 

The conditions for the economy equilibrium is
\begin{equation} \label{GrindEQ__4_} 
\sum _{k=1}^{M} c_{sk}^{1} \frac{D_{k} (p)}{\sum _{s=1}^{n} c_{sk}^{1} p_{s} } \le \sum _{k,j=1}^{M} e_{kj}^{s} =\psi _{s}^{1} ,\quad s=\overline{1,n}. 
\end{equation} 

As the statistical data are given in the cost form, it is convenient to rewrite the set of inequalities \eqref{GrindEQ__4_} in the cost form too
\begin{equation} \label{GrindEQ__5_} 
\sum _{k=1}^{M} c_{sk} \frac{D_{k} }{\sum _{s=1}^{n} c_{sk} } \le \psi _{s} ,\quad s=\overline{1,n}, 
\end{equation} 

where $c_{sk} =p_{s} c_{sk}^{1} ,\; \psi _{s} =p_{s} \psi _{s}^{1} ,\; D_{k} =\sum _{j=1}^{M} \sum _{s=1}^{n} p_{s} e_{kj}^{s} .$ We call the vector $c_{k} =\{ c_{sk} \} _{s=1}^{n} ,\; k=\overline{1,M},$ the demand vector of the $k$-th country and the vector $b_{k} =\{ b_{sk} \} _{s=1}^{n} ,\; k=\overline{1,M},$ the supply vector of the $k$-th country.

\textit{Definition 2}\textbf{. }We say the exchange by $n$ types of goods in cost form between $M$countries is at the equilibrium state if the set of inequalities \eqref{GrindEQ__5_} are true. The price vector $p=\{ p_{1} ,\ldots ,p_{n} \} $ under which the set of inequalities \eqref{GrindEQ__5_} are true is called the equilibrium price vector.

In what follows, we will use the denotations from [1]. So, we put $\sum _{j=1}^{M} p_{s} e_{kj}^{s} =b_{sk} ,\; s=\overline{1,n},\; k=\overline{1,M}$ and introduce the property vector of the $k$-th country $b_{k} =\{ b_{sk} \} _{s=1}^{n} ,\; k=\overline{1,n}.$ In these denotations the equilibrium state is written as
\begin{equation} \label{GrindEQ__6_} 
\sum _{k=1}^{M} c_{sk} \frac{D_{k} }{\sum _{s=1}^{n} c_{sk} } \le \psi _{s} ,\quad s=\overline{1,n}, 
\end{equation} 

where $\psi _{s} =\sum _{k=1}^{M} b_{sk} ,\; D_{k} =\sum _{s=1}^{n} b_{sk} .$ As the set of inequalities \eqref{GrindEQ__6_} may be not satisfied by the price vector $p=\{ p_{1} ,\ldots ,p_{n} \} ,$ we introduce the relative price vector $p_{0} =\{ p_{1}^{0} ,\ldots ,p_{n}^{0} \} ,$ to provide the equilibrium in the exchange model
\begin{equation} \label{GrindEQ__7_} 
\sum _{k=1}^{M} c_{sk} \frac{D_{k} (p_{0} )}{\sum _{s=1}^{n} p_{s}^{0} c_{sk} } \le \psi _{s} ,\quad s=\overline{1,n}, 
\end{equation} 

where $\psi _{s} =\sum _{k=1}^{M} b_{sk} ,\; D_{k} (p_{0} )=\sum _{s=1}^{n} p_{s}^{0} b_{sk} .$ It is evident that if the relative equilibrium price vector $p_{0} =\{ p_{1}^{0} ,\ldots ,p_{n}^{0} \} ,$ satisfying the set of inequalities \eqref{GrindEQ__7_} is such that $p_{i}^{0} =1,\; i=\overline{1,n},$ then the price vector $p=\{ p_{1} ,\ldots ,p_{n} \} ,$ is an equilibrium price vector. Introduced relative equilibrium vector $p_{0} =\{ p_{1}^{0} ,\ldots ,p_{n}^{0} \} $ characterizes the deviation of the exchange model from the equilibrium state.

What means that the exchange of $n$ types of goods in the cost form between $M$ countries is in an equilibrium state and why it is important to investigate the equilibrium states? First, let us note that if for every country export-import balance is equal zero, then the relative equilibrium vector $p_{0} =\{ p_{1}^{0} ,\ldots ,p_{n}^{0} \} $ is such that $p_{i}^{0} =1,\; i=\overline{1,n},$ and the price vector $p=\{ p_{1} ,\ldots ,p_{n} \} ,$ is an equilibrium price vector, moreover, the set of inequalities \eqref{GrindEQ__7_} becomes the set of equalities. This case is ideal one in the reality. We describe the factors that can violate the ideal equilibrium in the exchange model. In the exchange model, there exist equilibrium states differing from this ideal case. As every country seeks to protect its economy from goods flow from outside, it establishes the customs-tariffs. This leads to the violation of zero export-import balances between the countries. Another factor influencing the violation of zero export-import balance is the state in which the country needs more import than export due to the internal state of economy and vice versa when the country needs more export than import. Such deviations in the export-import balance can lead to the equilibrium states that are far from the ideal equilibrium state. The quality of such equilibrium states can be very different. The main aim is to investigate these equilibrium states. The deformation of the ideal equilibrium state in the exchange model that can arise may also lead to the recession in the world economy. The above established arguments are very important to investigate these deformed equilibrium states.

\section{International Trade of the 8Countries}


In this chapter we investigate the trade of 8 countriesbetween themselves.Our analysis is based on statistical data presented in[3,4].It is convenient to number them as:1. Canada, 2. China, 3. Germany, 4. France, 5. the United Kingdom,6. Italy7. Japan, 8. The United States.

These countries trade goods among themselves: 1. Animal, 2. Vegetable, 3. FoodProd, 4. Minerals, 5. Fuels, 6. Chemicals, 7. PlastiRub, 8. HidesSkin, 9. Wood, 10. TextCloth, 11. Footwear, 12. StoneGlas, 13. Metals, 14. MachElec, 15. Transport, 16. Miscellan.

During 2020 - 2022, trade relations between 8 countries were in non equilibrium states. The equilibrium state existed in each of the studied years. Each of these equilibrium states was far from ideal equilibrium. Each of the equilibrium states turned out to be highly degenerate. The degeneracy multiplicity was equal 15. An important concept of a generalized equilibrium price vector is introduced defined as a solution to a degenerate system of equations with real consumption. Using the concept of a generalized equilibrium vector, a recession level parameter is introduced. This parameter is a characteristic of the stability for the international exchange currency. The greater its value, the weaker the international exchange currency.

\section{ Partial Analysis of the International Trade of 8 Countries}


Below we presentan analysis in the form of diagrams of the parts of the demand and supply of the k-th country for the goods exchanged by the 8 countries. The same analysis is given in the form of diagrams of the supply and demand parts 8 countries for the k-th type of goods. Our analysis is based on statistical data [14,15].

Figures 1-3 show the export parts of each of the 8 countries.Below wegive an analysis of the supplyshares to the common market of the 8 most developed countries.In 2020, the share of China in the common market of goods exchanged by 8 countries is 0.28, the United States is 0.169, and Germany is 0.16.Already in 2021, the share of China's supply of goods is 0.291, while the supply of goods by the United States decreases to 0.165, and Japan decreases to 0.095. It should be noted that the share of Canada increased to 0.108.The trend is almost maintained in 2021. It is true that the share of China is becoming a little smaller and is 0.281, but the share of the USA is slightly increasing and is 0.176, and the share of Canada is also increasing and is 0.123. Note the decrease in the share of Germany, which is 0.155, and the share of Japan, which is 0.085.

The general conclusion is that there is a significant increase in China's presence on the market of 8 countries at the expense of such leading economies as the United States, Japan, and Germany.

Figures 4-6 show the dynamics of imports on the market of 8 countries.

\noindent \includegraphics*[width=3.26in, height=2.54in, trim=0.14in 0.05in 0.29in 0.15in]{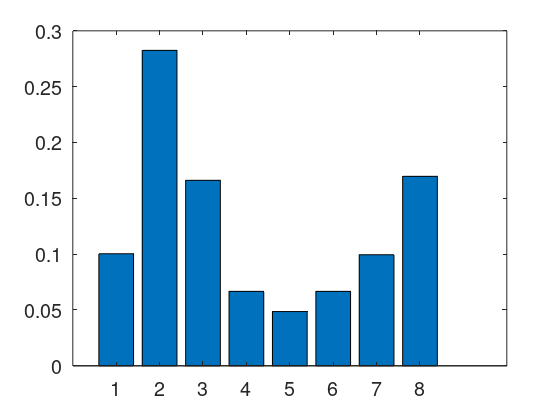}

\noindent \textbf{\textit{Figure 1.}}\textit{The part of supply of goods by the k-th country for goods of the eight. 2020.}

\noindent 1.Canada: 0.10, 2.China: 0.28, 3.Germany: 0.16, 4.France: 0.06, 5.The United Kingdom: 0.04, 6.Italy: 0.06, 7.Japan: 0.1, 8.The United States: 0.169.

\noindent \includegraphics*[width=3.28in, height=2.54in, trim=0.13in 0.07in 0.32in 0.13in]{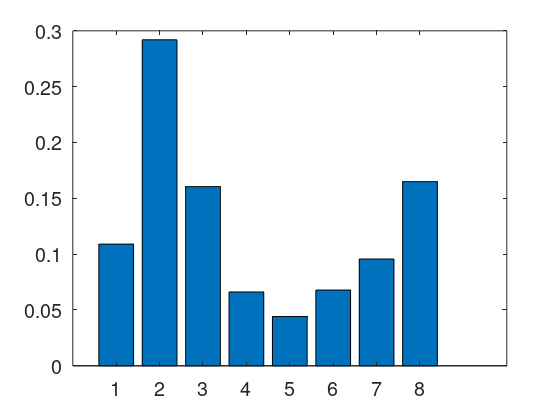}

\noindent \textbf{\textit{Figure 2.}}\textit{The part of supply of goods by the k-th country for goods of the eight. 2021.}

\noindent 1.Canada: 0.108, 2.China: 0.291, 3.Germany: 0.16, 4.France: 0.066, 5.The United Kingdom: 0.044, 6.Italy: 0.067, 7.Japan: 0.095, 8.The United States: 0.165.

\noindent \includegraphics*[width=3.28in, height=2.53in, trim=0.14in 0.07in 0.31in 0.13in]{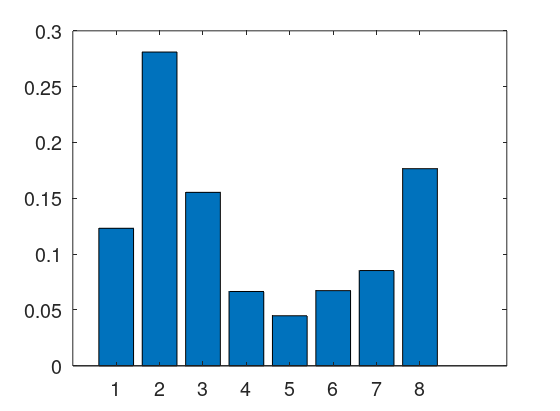}

\noindent \textbf{\textit{Figure 3.}}\textit{The part of supply of goods by the k-th country for goods of the eight. 2022.}

\noindent 1.Canada: 0.123, 2.China: 0.281,3. Germany: 0.155, 4.France: 0.066, 5.The United Kingdom: 0.044, 6.Italy: 0.067, 7.Japan: 0.085, 8.The United States: 0.176.

\noindent \includegraphics*[width=3.28in, height=2.55in, trim=0.14in 0.08in 0.31in 0.15in]{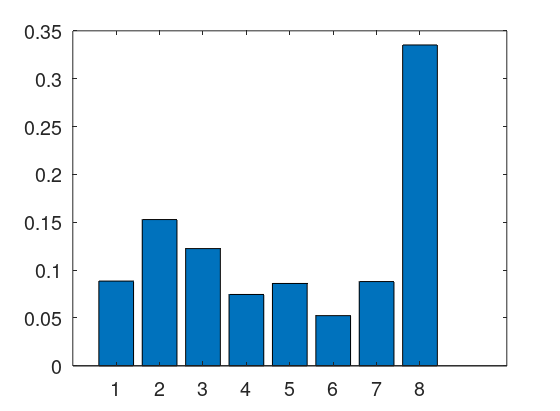}

\noindent \textbf{\textit{Figure 4.}}\textit{The part of demand of goods by the k-th country for goods of the eight. 2020.}

\noindent 1.Canada: 0.088, 2.China: 0.152, 3.Germany: 0.122, 4.France: 0.074, 5.The United Kingdom: 0.086, 6.Italy: 0.052, 7.Japan: 0.088, 8.The United States: 0.335.

\noindent \includegraphics*[width=3.25in, height=2.51in, trim=0.13in 0.08in 0.31in 0.15in]{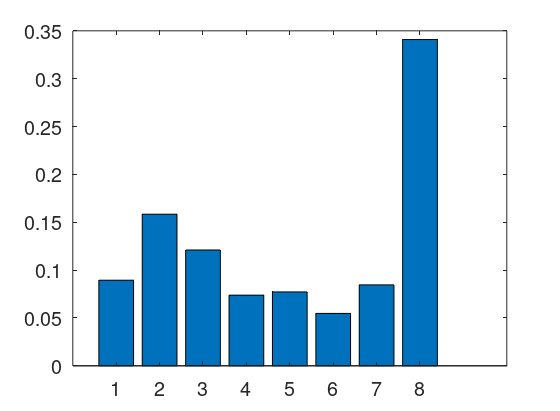}

\noindent \textbf{\textit{Figure 5.}}\textit{The part of demand of goods by the k-th country for goods of the eight. 2021.}

\noindent 1.Canada: 0.089, 2.China: 0.158, 3.Germany: 0.1209, 4.France: 0.0737, 5.The United Kingdom: 0.0771, 6.Italy: 0.054, 7.Japan: 0.084, 8.The United States: 0.3409.

\noindent \includegraphics*[width=3.27in, height=2.52in, trim=0.14in 0.07in 0.32in 0.14in]{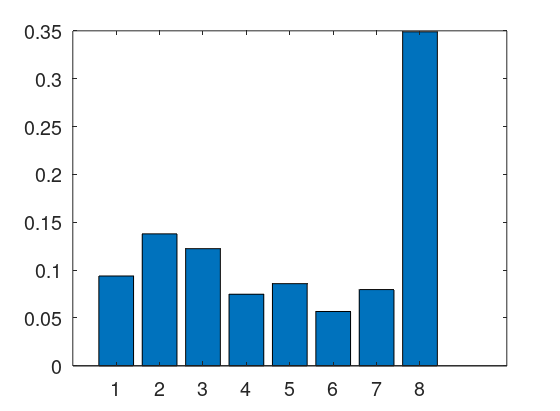}

\noindent \textbf{\textit{Figure 6.}}\textit{The part of demand of goods by the k-th country for goods of the eight. 2022.}

\noindent 1.Canada: 0.093, 2. China: 0.137, 3.Germany: 0.122, 4.France: 0.074, 5.The United Kingdom: 0.085, 6.Italy: 0.056, 7.Japan: 0.079, 8.The United States: 0.348.

\noindent \includegraphics*[width=3.28in, height=2.51in, trim=0.15in 0.06in 0.27in 0.14in]{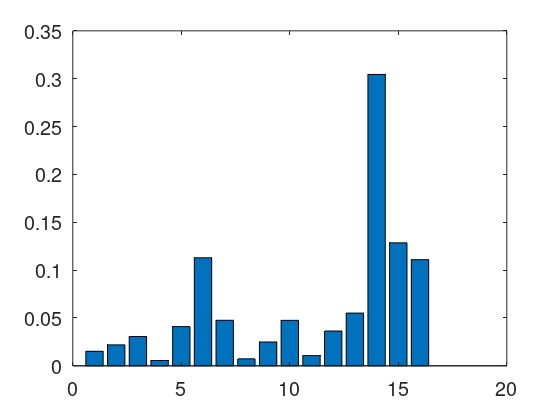}

\noindent \textbf{\textit{Figure 7. }}\textit{The part of demand for the k-th type of goods by all 8 country. 2020.}

\noindent 1. Animal: 0.015 2. Vegetable: 0.021 3. FoodProd: 0.030 4. Minerals: 0.005 5. Fuels: 0.040, 6. Chemicals: 0.112, 7. PlastiRub: 0.047, 8. HidesSkin: 0.007, 9. Wood: 0.024, 10.TextCloth: 0.047, 11. Footwear: 0.010, 12. StoneGlas: 0.036, 13. Metals: 0.055, 14. MachElec: 0.304, 15. Transport: 0.128, 16. Miscellan: 0.1108.

\noindent \includegraphics*[width=3.25in, height=2.48in, trim=0.13in 0.08in 0.26in 0.14in]{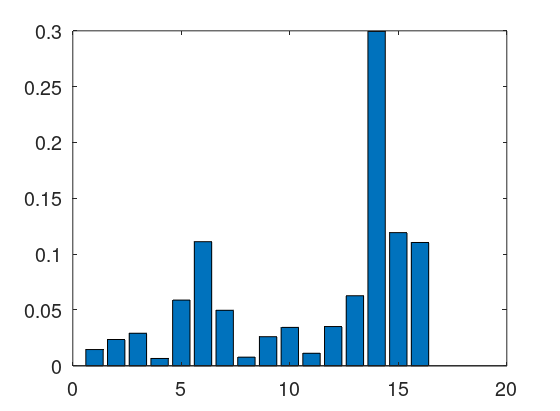}

\noindent \textbf{\textit{Figure 8. }}\textit{The part of demand for the k-th type of goods by all 8 country. 2021.}

\noindent 1. Animal: 0.014 2. Vegetable: 0.023 3. FoodProd: 0.029 4. Minerals: 0.006 5. Fuels: 0.058, 6. Chemicals: 0.111, 7. PlastiRub: 0.049, 8. HidesSkin: 0.007, 9. Wood: 0.026, 10.TextCloth: 0.034, 11. Footwear: 0.011, 12. StoneGlas: 0.035, 13. Metals: 0.062,14. MachElec: 0.299, 15. Transport: 0.119, 16. Miscellan: 0.110.

\noindent \includegraphics*[width=3.25in, height=2.51in, trim=0.12in 0.07in 0.27in 0.13in]{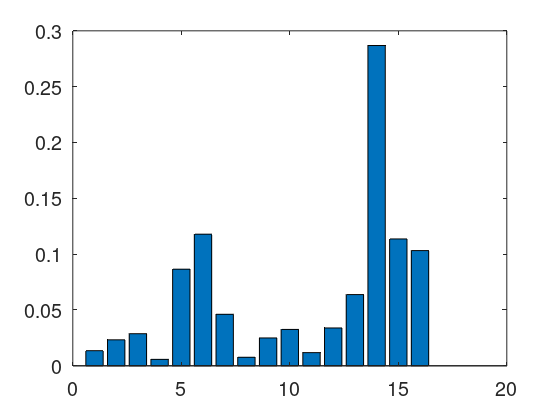}

\noindent \textbf{\textit{Figure 9. }}\textit{The part of demand for the k-th type of goods by all 8 country. 2022.}

\noindent 1. Animal: 0.013 2. Vegetable: 0.023 3. FoodProd: 0.028 4. Minerals: 0.005 5. Fuels: 0.086, 6. Chemicals: 0.117, 7. PlastiRub: 0.046, 8. HidesSkin: 0.007, 9. Wood: 0.024, 10.TextCloth: 0.032, 11. Footwear: 0.011, 12. StoneGlas: 0.033, 13. Metals: 0.063,14. MachElec: 0.286, 15. Transport: 0.114, 16. Miscellan: 0.103.

\noindent \includegraphics*[width=3.26in, height=2.46in, trim=0.12in 0.07in 0.25in 0.14in]{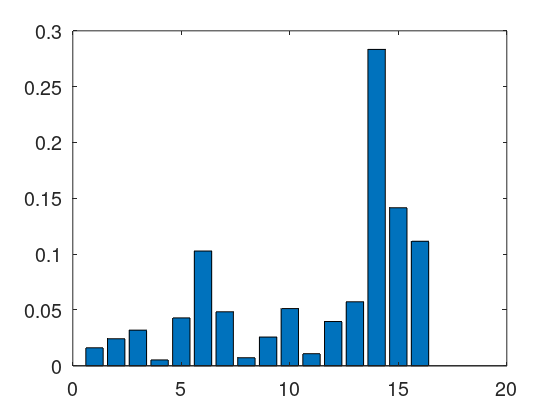}

\noindent \textbf{\textit{Figure 10. }}\textit{The part of supply of the k-th type of goods by all 8 country. 2020.}

\noindent 1. Animal: 0.016 2. Vegetable: 0.024 3. FoodProd: 0.031 4. Minerals: 0.005 5. Fuels: 0.042,6. Chemicals: 0.102, 7. PlastiRub: 0.048, 8. HidesSkin: 0.007, 9. Wood: 0.025, 10.TextCloth: 0.051, 11. Footwear: 0.010, 12. StoneGlas: 0.039, 13. Metals: 0.057,14. MachElec: 0.283, 15. Transport: 0.141, 16. Miscellan: 0.111.

\noindent \includegraphics*[width=3.25in, height=2.50in, trim=0.13in 0.07in 0.25in 0.14in]{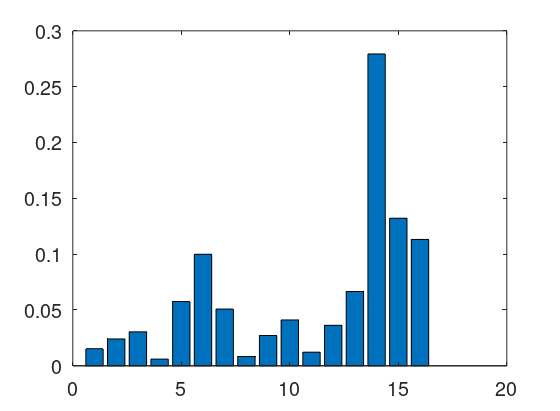}

\noindent \textbf{\textit{Figure 11. }}\textit{The part of supply of the k-th type of goods by all 8 country. 2021.}

\noindent 1.Animal: 0.015, 2.Vegetable: 0.024, 3.FoodProd: 0.030, 4.Minerals: 0.005, 5. Fuels: 0.05,6. Chemicals: 0.099, 7. PlastiRub: 0.05, 8. HidesSkin: 0.008, 9. Wood: 0.027,10.TextCloth: 0.041, 11. Footwear: 0.012, 12. StoneGlas: 0.036, 13. Metals: 0.066,14. MachElec: 0.279, 15. Transport: 0.132, 16. Miscellan: 0.113.

\noindent \includegraphics*[width=3.25in, height=2.51in, trim=0.13in 0.07in 0.26in 0.13in]{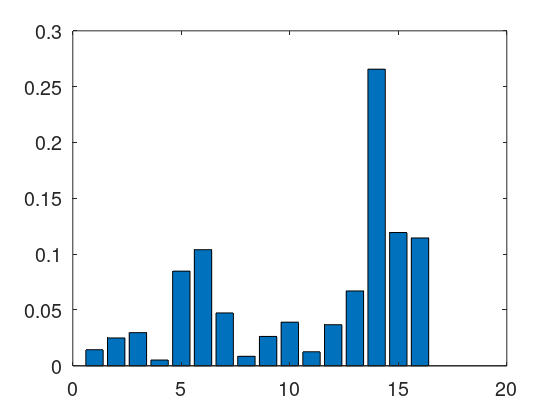}

\noindent \textbf{\textit{Figure 12. }}\textit{The part of supply of the k-th type of goods by all 8 country. 2022.}

\noindent 1.Animal: 0.014, 2.Vegetable: 0.024, 3.FoodProd: 0.029, 4.Minerals: 0.005, 5. Fuels: 0.084,6. Chemicals: 0.103, 7. PlastiRub: 0.047, 8. HidesSkin: 0.008, 9. Wood: 0.026,10.TextCloth: 0.039, 11. Footwear: 0.012, 12. StoneGlas: 0.036, 13. Metals: 0.067,14. MachElec: 0.265, 15. Transport: 0.119, 16. Miscellan: 0.114.

\noindent The share of China's imports from the remaining 7 countries in 2020 is 0.152, the share of the United States is 0.335, and the share of Germany is 0.122.In 2021, we see an increase in the share of China's imports to 0.158, but there is a drop in the share of US imports to 0.3409 and a decrease in the share of Germany to 0.1209.In 2021, the share of China's imports increases to 0.158, the US maintains the same share as in the previous year.It should be noted that the shares of imports byFrance, Great Britain and Italy are higher than the shares of exports.

In 2022, the situation changes dramatically, the share of China's imports drops sharply to 0.137, while the share of the United States rises sharply to 0.348. The share of Germany's imports is also increasing, which is 0.122.

\noindent The main conclusion is that 2022 is a turning point for China's imports from the rest of the world's most developed countries.

The dynamics of changes in the shares of demand for goods exchanged by the most developed countries of the world are presented below.Its characteristic feature is the growth in demand for energy products, chemical products and metal products, which indicates the growth of production during the years under review.The share of demand for engineering and transport products is falling.A drop in the share of demand for food products is observed.
\[2020.\] 

3. FoodProd: 0.030

5. Fuels: 0.040,

6. Chemicals: 0.112,

7. PlastiRub: 0.047,

10. TextCloth: 0.047,

12. StoneGlas: 0.036,

13. Metals: 0.055,

14. MachElec: 0.304,

15. Transport: 0.128,

16. Miscellan: 0.1108.
\[2021.\] 

3. FoodProd: 0.029

5. Fuels: 0.058,

6. Chemicals: 0.111,

7. PlastiRub: 0.049,

10. TextCloth: 0.034,

12. StoneGlas: 0.035,

13. Metals: 0.062,

14. MachElec: 0.299,

15. Transport: 0.119,

16. Miscellan: 0.110.
\[2022.\] 

3. FoodProd: 0.028,

5. Fuels: 0.086,

6. Chemicals: 0.117,

7. PlastiRub: 0.046,

10. TextCloth: 0.032,

12. StoneGlas: 0.033,

13. Metals: 0.063,

14. MachElec: 0.286,

15. Transport: 0.114,

16. Miscellan: 0.103.

\noindent The dynamics of supply shares in the market of the 8 most developed economies of the world are presented below.There is a decrease in the share of food products, an increase in the share of the supply of energy carriers, and an increase in the supply of metal products.There is a drop in the supply share of electrical engineering, and the share of transportation is decreasing.
\[2020.\] 

3. FoodProd: 0.031,

5. Fuels: 0.042,

6. Chemicals: 0.102,

7. PlastiRub: 0.048,

10. TextCloth: 0.051,

12. StoneGlas: 0.039,

13. Metals: 0.057,

14. MachElec: 0.283,

15. Transport: 0.141,

16. Miscellan: 0.111.
\[2021.\] 

3.FoodProd: 0.030,

5. Fuels: 0.05,

6. Chemicals: 0.099,

7. PlastiRub: 0.05,

10.TextCloth: 0.041,

12. StoneGlas: 0.036,

13. Metals: 0.066,

14. MachElec: 0.279,

15. Transport: 0.132,

16. Miscellan: 0.113.
\[2022\] 

3.FoodProd: 0.029,

5. Fuels: 0.084,

6. Chemicals: 0.103,

7. PlastiRub: 0.047,

10.TextCloth: 0.039,

12. StoneGlas: 0.036,

13. Metals: 0.067,

14. MachElec: 0.265,

15. Transport: 0.119,

16. Miscellan: 0.114.

\section{Conclusions}

The work presents the analysis of international trade of the 8 most developed economies of the world. It is shown that the trade between them was not in a state of economic equilibrium. The relative equilibrium price vector turned out to be highly degenerate. A detailed analysis of the market shares occupied by each country is given.
\section{Abbreviations}

\begin{tabular}{|p{0.6in}|p{4.1in}|} \hline 
G7 & The Group of Seven is a group of the seven advanced economies in the world: Canada, France, Germany, Italy, Japan, the United Kingdom and the United States. \\ \hline 
 &  \\ \hline 
 G20 & The G20 (or Group of Twenty) is an international forum for the governments and central bank governors    from 19 countries and the European Union (EU). \\ \hline 
\end{tabular}

\textbf{}

\section{Funding}

This work is partially supported by the Fundamental Research Program of the Department of Physics and Astronomy of the National Academy of Sciences of Ukraine "Building and researching financial market models using the methods of nonlinear statistical physics and the physics of nonlinear phenomena N 0123U100362."


\noindent The authors declare no conflicts of interest.

\vskip 15mm

\noindent \textbf{ \Large{References}}
\vskip 3mm

\noindent [1] Gonchar, N.S., Dovzhyk, O.P., Zhokhin, A.S., Kozyrski, W.H. and Makhort, A.P. International Trade and Global Economy. Modern Economy, 2022, 13. 901-943.https://doi.org/10.4236/me.2022.136049

\noindent [2]  Gonchar, N. S. Mathematical foundations of information economics.Bogolyubov Institute for Theoretical Physics, Kiev, 2008,  468p.

\noindent [3] Gonchar, N.S. Economy Function in the Mode of Sustainable Development. Advances in Pure Mathematics, 2024, 14, 242-282. https://doi.org/10.4236/apm.2024.144015

\noindent 

\noindent [4] Gonchar, N.S. Economy Equilibrium and Sustainable Development. Advances in Pure Mathematics,2023,13, 316-346. https://doi.org/10.4236/apm.2023.136022

\noindent 

\noindent [5] Gonchar, N.S. Mathematical Foundations of Sustainable Economy Development. Advances in Pure Mathematics, 2023, 13, 369-401. https://doi.org/10.4236/apm.2023.136024

\noindent 

\noindent [6] Gonchar, N.S. and Dovzhyk, O.P. On the Sustainable Economy Development of Some European Countries. Journal of Modern Economy, 2022, 5, 1-14. https://doi.org/10.28933/jme-2021-12-0505

\noindent 

\noindent [7] Gonchar, N. S. Mode of sustainable economic development, arXiv:2405.09984 [econ.GN], 2024, https://doi.org/10.48550/arXiv.2405.09984

\noindent 

\noindent [8] Yali Chen, Zheng Xiang Application of economic model optimization algorithm for international trade based on bigdata technology, 2023, 14~p. https://doi.org/10.2478/10.2478/amns.2023.2.00041

\noindent [9]  Rahman, A.A.  The Basic Laws of Trade: Reconstructing the Theory of International Trade. 2022,  GCI Working Paper No 2-2022, 49~p. https://doi.org/10.33774/coe-2022-qjrf5-v6

\noindent [10] Piotr Rubaj International Tradeas a Key Factor for Sustanable Economic Development. European Research Studies Journal Vol XXV, Issue 4, 2022, p. 195--206, doi: 10.35808/ersj/3075

\noindent [11] Lewkowicz, J. Thoughtson the Political Economy of International Trade. Central European Economic Journal, 2024, 11\eqref{GrindEQ__58_}, 33--41. doi: 10.2478/ceej-2024-0004

\noindent [12] Zdzis{\l}aw W. Pu\'{s}lecki Rebuilding Of International Trade And Investment , ISRG Journal of Economics, Business, \& management, 2024, Vol.II, Issue II, p. 56--65. doi: 10.5281/zenodo.10864865

\noindent [13] Wentao Cui, Ziyu Li Reasons for International Trade Barriers and Suggestions to Eliminate Trade Barriers. Proc. 2nd International Conference on Financial Technology and Business Analysis, 2023, p.141--151. doi: 10.54254/2754-1169/97/20231563

\noindent [14] World Integrated Trade Solution. WITS. Available from: http://wits.worldbank.org/(accessed 6 October 2022).

\noindent [15] The Organization for Economic Co-operation and Development. OECD. Available from: https://www.oecd.org/accessed 6 October 2022).

\noindent 

\noindent 

\vskip 5mm

\noindent \textbf{China and G7 in the Current Context of the World Trading}

\noindent \textbf{Nicholas Simon Gonchar, Olena Petrivna Dovzhyk, Anatoly Sergiyovych Zhokhin,}

\noindent \textbf{Wolodymyr Hlib Kozyrski, Andrii Pylypovych Makhort}

\noindent Mathematical Modelling Labaratory of Synergetics, Department, Bogolyubov Institute for Theoretical Physics of NAS of Ukraine, Kyiv, Ukraine.

\noindent 

\vskip 5mm

\noindent https://orcid.org/0000-0003-0954-8948 (N.S.Gonchar), https://orcid.org/0000-0003-4098-1192 (O.P.Dovzhyk)

\noindent https://orcid.org/0000-0001-7826-6608 (A.S.Zhokhin), https://orcid.org/0000-0002-8969-3928 (W.H.Kozyrski)

\noindent https://orcid.org/0000-0002-8877-5884 (A.P.Makhort)

\end{document}